\documentclass[nameyear]{elsart}
\usepackage{epsfig}
\usepackage{amssymb}
\usepackage{natbib}
\usepackage{color}

\newcommand{\Msun}{\mbox{$M_\odot$}}

\newcommand{\apj}{ApJ}
\newcommand{\aap}{A\&A}
\newcommand{\apjs}{ApJS}
\newcommand{\apjl}{ApJ Lettr}
\newcommand{\prc}{Phys. Rev. C}
\newcommand{\physrep}{Physics Reports}
\def\be{\begin{eqnarray}}
\def\ee{\end{eqnarray}}
\def\lsim{\mathrel{\rlap{\lower3pt\hbox{\hskip1pt$\sim$}}
     \raise1pt\hbox{$<$}}} 
\def\gsim{\mathrel{\rlap{\lower3pt\hbox{\hskip1pt$\sim$}}
     \raise1pt\hbox{$>$}}} 

\newcommand{\lSect}[1]{{\label{sec:#1}}}
\newcommand{\lFig}[1]{{\label{fig:#1}}}

\newcommand{\PAN}[1]{{{#1}}}

\newcommand{\FIGFF}[2]{{\ref{fig:#2}\PAN{#1}}}

\newcommand{\FIG}[2]{{Fig.~\FIGFF{#1}{#2}}}
\newcommand{\Fig}[1]{{\FIG{}{#1}}}

\newcommand{\Sectff}[1]{{\ref{sec:#1}}}
\newcommand{\Sect}[1]{{\S\Sectff{#1}}}

\begin{document}

\runauthor{Woosley \& Heger}

\begin{frontmatter}
\title{Nucleosynthesis and Remnants in Massive Stars of Solar Metallicity}

\author[ucsc]{S. E. Woosley}
\author[ucsc,lanl]{A. Heger}

\address[ucsc]{
Department of Astronomy and Astrophysics,
UCSC, Santa Cruz  CA  95064\\ (E-mail: woosley@ucolick.org) }
\address[lanl]{Theoretical Astrophysics Group, T-6, MS B227, Los Alamos National Laboratory,
Los Alamos  NM 87545}

\begin{abstract}
Hans Bethe contributed in many ways to our understanding of the
supernovae that happen in massive stars, but, to this day, a first
principles model of how the explosion is energized is
lacking. Nevertheless, a quantitative theory of nucleosynthesis is
possible. We present a survey of the nucleosynthesis that occurs in 32
stars of solar metallicity in the mass range 12 to 120 \Msun.  The
most recent set of solar abundances, opacities, mass loss rates, and
current estimates of nuclear reaction rates are employed. Restrictions
on the mass cut and explosion energy of the supernovae based upon
nucleosynthesis, measured neutron star masses, and light curves are
discussed and applied. The nucleosynthetic results, when integrated
over a Salpeter initial mass function (IMF), agree quite well with
what is seen in the sun. We discuss in some detail the production of
the long lived radioactivities, $^{26}$Al and $^{60}$Fe, and why
recent model-based estimates of the ratio $^{60}$Fe/$^{26}$Al are
overly large compared with what satellites have observed. A major
source of the discrepancy is the uncertain nuclear cross sections for
the creation and destruction of these unstable isotopes.
\end{abstract}

\end{frontmatter}


\section{Introduction}
\lSect{intro}

Starting in the late 1970's, with the encouragement of his good friend
Gerry Brown, Hans Bethe became interested in applying his expertise in
nuclear physics to one of the more vexing problems in modern
astrophysics - how massive stars die as supernovae. The problem is
difficult for a variety of reasons. The iron core of a massive star
collapses to a neutron star (or sometimes a black hole) and, somehow,
some fraction of that remnant's binding energy is converted into
outwards kinetic energy in the overlying star.  The favored model, now
as then, says that the binding energy of the neutron star is radiated
as neutrinos, a fraction of which deposit their energy in the matter
above the neutron star causing it to expand and explode \citep{Col66}.

When Hans began to work on the problem, supernova models were not
giving explosions. Moreover, the physics was very uncertain with
bounce densities ranging from 10$^{13}$ g cm$^{-3}$ to 10$^{15}$ g
cm$^{-3}$ \citep[e.g.,][]{Wil78}. The nuclear equation of state was
particularly uncertain. A major breakthrough was the work by
\cite{BBAL} who showed that the heat capacity of the nuclear bound
states was much larger than previously believed \citep{Fow78}. In
fact, the mean excitation energy was $E_x \sim (A/8) (kT)^2$ and the
partition function associated with all these states was exponentially
huge, $G \sim \exp((A/8) \ (kT))$. Consequently, nuclear equilibrium
favored bound nuclei which remained abundant, increasing their average
mass, until they touched and merged near nuclear density. Bounce
occurred at super-nuclear density on the hard core, repulsive
component of the strong force (not thermal pressure as some
calculations claimed) and was at low entropy. The general idea of
entropy as an important variable in core collapse came from Hans, who
liked to remark that though the bounce was thermally very hot, in
terms of entropy it was as ordered as ice.

During the next 20 years, Hans made many other lasting contributions
to supernova theory, including the currently favored ``delayed''
neutrino transport paradigm in which convection plays a major role
\citep{Bet85}\footnote{The first calculation to show the revival of
the shock by neutrino heating was carried out by Wilson alone in 1982
\citep{Wil85}, but analysis of the calculation and the first refereed
publication was by Bethe and Wilson. For a time, Hans also embraced
the idea of ``prompt explosions'' \citep{Bar87}, explosions in which
neutrino transport played no constructive role and the explosion was
due to a hydrodynamical ``bounce''. He gave up the idea after detailed
calculations showed that neutrino losses and photodisintegration
killed the prompt shock.} Hans also introduced the ideas of a ``gain
radius'', where neutrino heating first exceeds neutrino losses, and of
``net ram'', the momentum of the accretion flux that must be overcome
to get the shock to move out. He excelled in simple analytic models
for the physics of core collapse, and brought a much needed physically
intuitive understanding of a subject that had hitherto been largely
numerical \citep[e.g.,][]{Bet90}.

Because of his historical interests in stellar structure and nuclear
physics, Hans was also interested in the presupernova evolution of
massive stars and in nucleosynthesis. During his visits to Santa Cruz
and by mail, we had many discussions on the progenitor of SN 1987A,
the physics of supernova light curves, the nature of the ``reverse
shock'', explosive nucleosynthesis, and on the $r$-process. Thus it is
to his memory that this paper is dedicated.

To this day, we still don't know exactly how massive stars explode
\citep{Woo05}, so the parameterization of the explosion is discussed
in \Sect{expl}. The key nuclear reaction rates and other uncertain
aspects of the presupernova evolution are described in
\Sect{uncertainty}.  In \Sect{nucleo}, our principal nucleosynthetic
results are presented, and in \Sect{fe60}, we conclude with a
discussion of two key species of interest in $\gamma$-line astronomy,
$^{26}$Al and $^{60}$Fe.

Throughout this paper and in the future, we employ a unit of energy,
the ``Bethe'', abbreviated ``B'', equal to $1.0 \times 10^{51}$
erg. Gerry Brown introduced, and Hans and Gerry both promoted the use
of an alternate term, ``foe'', frequently found in the supernova
literature to stand for 10$^{51}$ erg, but in deference to Hans'
contributions to the field, we follow the convention suggested by
\cite{Wei06}.

\section{Uncertainties in the Presupernova Evolution}
\lSect{uncertainty}

\subsection{Critical Reaction Rates}
\lSect{c12ag}

The key uncertain reaction rate affecting both the structure of and
nucleosynthesis in massive stars remains
$^{12}$C($\alpha,\gamma)^{16}$O, despite over 30 years of painstaking
laboratory investigation \citep[e.g.,][]{Dye74}. The experimental
situation was recently reviewed by \citet{Buc05}, who recommends S(300
keV) = 102 - 198 keV\,b with a best value of 145 keV\,b.  Based upon
nucleosynthesis considerations, \citet{Woo93} estimated an S-factor of
$\sim$170 keV\,b, which remains within the experimental range
today. More precisely, Weaver and Woosley suggested a rate $1.7 \pm
0.5$ times that of \cite{Cau88}, which would be 120 - 220 keV\,b, but
even at the time, the error bar was regarded as liberal. More
recently, \cite{Boy02} has revised the nucleosynthesis constraints
using more stellar models, a finer grid of
$^{12}$C($\alpha,\gamma)^{16}$O rates, finer stellar zoning, and other
improvements to the stellar model. Their results, shown in \Fig{fig1},
are in good agreement with the earlier calculations of Weaver and
Woosley, but give a narrower error bar and also make the sensitivity
of the results to this rate (variations of only 10\% matter) more
apparent. Because of the need to include a rate that is accurate
across a wide range of temperature, not just during helium burning,
the preferred rate is again expressed as a multiple of a published
rate fit, this time \cite{Buc96,Buc97}, which has S(300 keV) = 146
keV\,b. Boyes' best fit is about 1.2 times this, or 175 keV\,b and
{\sl a value of 1.2 times Buchmann (1996) was used in the present
study}. This is also consistent with recent measurements reported by
\cite{Ham05} that give a best value of 1.08 times Buchmann-1996 (i.e.,
162 $\pm 39$ keV\,b). Of course, one could argue that the
nucleosynthesis limit is also influenced by our uncertain model of
stellar convection \citep{Woo93}, in which case an experimental value
ultimately near 170 keV\,b would serve to validate the treatment of
convection in the code.

During the end of helium burning the $^{12}$C($\alpha,\gamma)^{16}$O
rate competes with the triple-alpha reaction rate, and hence the
uncertainty in that rate can have similar effects.  In test
calculations at $3\times10^8\,$K and 1000 and 2000 g/cm$^3$ we found
that a 10\,\% increase in the triple-alpha rate would have the same
consequence as an 8\,\% decrease in $^{12}$C($\alpha,\gamma)^{16}$O.
In a star, convection may change these results, though probably not
much. The $^{12}$C($\alpha,\gamma)^{16}$O rate would need to be known
better than about 10\,\% before the uncertainty in the triple
alpha rate, $\sim$12\,\% \citep{Tur07}, becomes a limiting factor.

\begin{figure}
\centerline{\epsfig{file=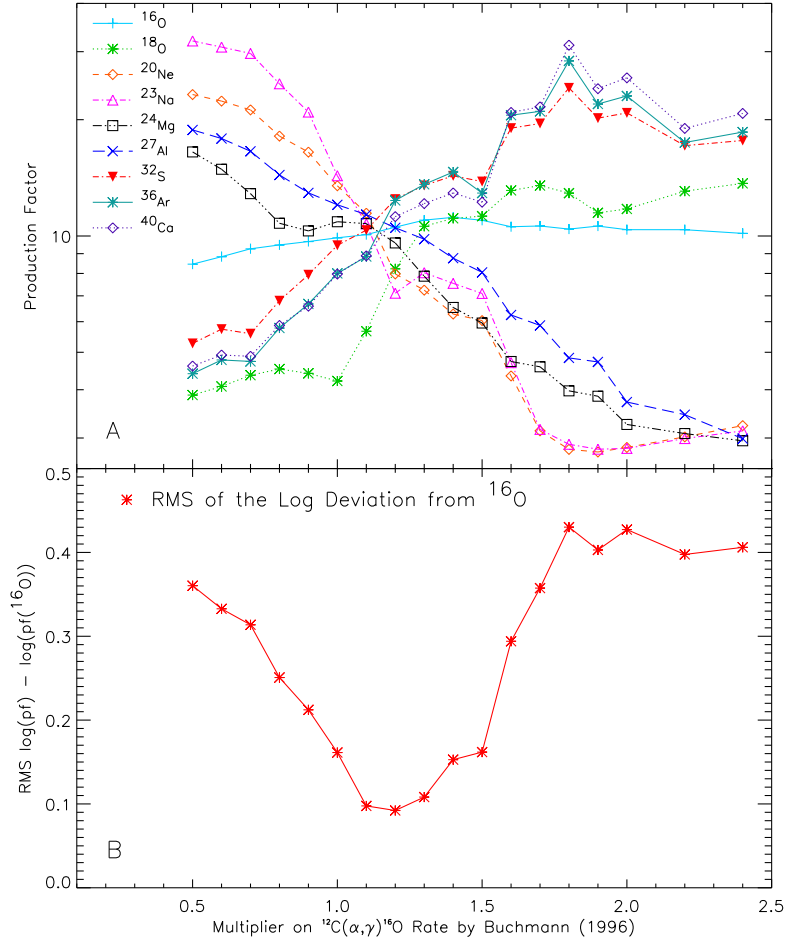,height=5in}}
\caption{Production factor factor of key elements for a set of solar
  metallicity stars folded with a Saltpeter birth function
  \citep{Boy02}.}
\lFig{fig1}
\end{figure}

The other uncertain reaction rate that affects the abundances of hosts
of nuclei, not just a few, is $^{22}$Ne($\alpha,$n)$^{25}$Mg, which,
along with $^{25}$Mg(n,$\gamma$)$^{26}$Mg, regulates the strength of
the $s$-process in massive stars. The rate used here is the
recommended value from from \cite{Jae01}. The reaction
$^{22}$Ne($\alpha,\gamma)^{26}$Mg competes with
$^{22}$Ne($\alpha,$n)$^{25}$Mg for the destruction of $^{22}$Ne and is
thus also of some importance for determining the strength of the
$s$-process. Here we use the lower bound for
$^{22}$Ne($\alpha,\gamma)^{26}$Mg estimated by \cite{Kap94}.  Other
choices of strong and weak reaction rates have been discussed by \cite{Woo02}. In
particular, except where otherwise noted, we use the Hauser-Feshbach
rates from \cite{Rau00} for reactions involving n, p, and $\alpha$ on
heavy nuclei that lack experimental determination. This is of some
relevance to the issue of $^{26}$Al and $^{60}$Fe production discussed
later in the paper (\Sect{nuc60}).

\subsection{Mass Loss}
\lSect{mdot}

Mass loss is known to be a powerful determinant in the evolution of
stars of nearly solar metallicity, and its omission was one of the
major shortcomings of the Woosley-Weaver 1995 survey \cite{WW95}. For
stars more massive than about 35 \Msun, mass loss is particularly
important since it not only removes the hydrogen envelope, but shrinks
the helium core appreciably. With current estimates of mass loss, a
100 \Msun\ Population I star ends its life as a star of only about 6
\Msun, composed of helium and heavier elements only and no hydrogen
left.  This is similar to the mass and composition of the core of a 20
\Msun\ star, and the explosion properties, remnant mass, and
nucleosynthesis are radically different from a 100 \Msun\ star that
had no mass loss (see \Sect{remnants}).

The mass loss prescription used here has also been discussed by
\cite{Woo02}. In particular, we use \cite{Nie90} for mass loss on the
main sequence and for red giants and \cite{WL99} for Wolf-Rayet stars
The latter is based on the mass loss rate by \cite{Bra97} fit to
observational data and divided by a factor of three to account for
clumping \cite{Ham98}.  The nucleosynthesis products carried away by
stellar wind are included in all yields reported in this paper.

\subsection{Convection and Rotation}
\lSect{conv}

The treatment of convective physics, including overshoot mixing and
semiconvection, follows the discussion in \cite{Woo88} and
\cite{Woo02}.  In particular, we use a semi-convective mixing
parameter, $\alpha$ = 0.1, which results in relatively fast mixing in
semiconvective regions.  Mixing was treated in a time-dependent,
mixing length formalism using the Ledoux criterion for instability.
The fast semiconvection contributes significantly to mixing in regions
that are stable to the Ledoux criterion but unstable to the
Schwarzschild criterion, however, the mixing is less than that of a
mere Schwarzschild mixing, taking into account the stabilizing
effects of composition gradients.

Rotation can have a significant effect on both the presupernova
evolution and the explosion mechanism. Here rotation was neglected,
which is to say it is assumed that the change in helium core mass and
dredge up of light isotopes due to rotationally induced mixing are
small, for moderately fast rotating stars, and that the ratio of
centrifugal force to gravity during the explosion is negligible.  All
these assumptions are questionable for rapidly rotating stars,
especially so in small fraction of massive stars that become gamma-ray
bursts \citep[e.g.,][]{Woo06}.

\subsection{Initial Abundances}
\lSect{solabn}

The evolution and nucleosynthesis of a massive star both are sensitive
to its initial composition. The total abundance of CNO affects the
efficiency of hydrogen burning and the opacity. The conversion of CNO
into $^{22}$Ne during helium burning determines the ``neutron
excess'', which affects the production of all nuclei with unequal
neutron and proton numbers. $^{22}$Ne also provides the free neutrons
necessary for the $s$-process during helium burning.  Finally, because
the yields of supernovae are traditionally normalized to $^{16}$O, any
change in the solar oxygen abundance affects the comparative ease with
which all other heavy elements are produced.

It is thus a major occurrence in nucleosynthesis theory when the solar
abundances, traditionally taken as representative of Population I
stars in our Galaxy, are modified.  Recent revisions to the solar
abundance set have been discussed by \cite{Lod03} and \cite{Asp04}. The
abundances of all isotopes of CNO have been reduced by amounts of
order 30\,\% compared with the standard Anders and Grevesse values
\citep{And89} of a few years ago. Here we use the \cite{Lod03} set
both as a starting composition, and also to normalize all computed
yields.

\subsection{Presupernova Models}
\lSect{presn}

Using the \textsc{Kepler} implicit hydrodynamics code \citep{Wea78}
and the physics specified above and in \cite{Woo02}, stars of solar
composition and various masses were evolved to the presupernova stage
- defined by a collapse velocity in the core of 1000
km\,s$^{-1}$.  Masses included in the study were 12 through 33 solar
masses in steps of 1 \Msun, plus stars of 35, 40, 45, 50, 55, 60,
70, 80, 100, and 120 \Msun\ - 32 stars altogether. A future survey
will use a much finer grid of masses, and the present work may be
regarded as a preliminary survey.

\section{Simulating the Explosion}
\lSect{expl}

As alluded to in the introduction, a robust description for how
massive stars explode as supernovae remains elusive and this must
surely affect our understanding of the origin of the
elements. It is worth separating out that part of the
nucleosynthesis that depends on the explosion mechanism from that
which does not, however.

Certainly isotopes produced in the vicinity of what is commonly known
as ``the mass cut'' are sensitive to conditions set up by the passage
of the shock. This includes the yields of species made by explosive
oxygen and silicon burning and in nuclear statistical
equilibrium. More quantitatively, these are the species made at
temperatures above $3 \times 10^9$ K and at radii less than about 7000
km, i.e., roughly the inner 1 to 2 solar masses of ejecta. Other
species made by hydrogen, helium, carbon, neon and oxygen burning in
hydrostatic equilibrium are not greatly affected (provided such
material escapes the star and does not fall into a black hole), nor is
the nucleosynthesis in the pre-explosive wind.  On the other hand, the
$r$-process and other species made in the neutrino-powered wind
\emph{are} quite sensitive to the explosion mechanism, and it is this
sensitivity that makes them excellent diagnostics of the event.

As we shall see though, even the ``explosive nucleosynthesis'' below
atomic mass 100 is not particularly sensitive to details of the
explosion, provided that the star blows up with a ``reasonable''
kinetic energy and the explosion is not grossly asymmetric. This is
basically because the shock conditions are determined by the
pre-explosive structure and some simple physics, $4 \pi R^3 a T^4/3$ =
explosion energy $\approx$ 1 B.

Here, as elsewhere, the explosion is parameterized by a piston at
constant Lagrangian mass coordinate that moves through the star with
some specified radial history \citep{WW95,Woo02}. The essential
parameters of the piston are its location in mass and the final
kinetic energy it imparts to the ejecta at infinity. Two different
choices of each are explored: a) piston mass at the edge of the iron
core or at the point where the dimensionless entropy $S/N_A k$ = 4.0;
and b) kinetic energies of 1.2 and 2.4 B. Thus for each mass, 4
explosion models were calculated for a total of 128 supernovae
simulated.

\begin{figure}
\vskip -1.0in
\centerline{\epsfig{file=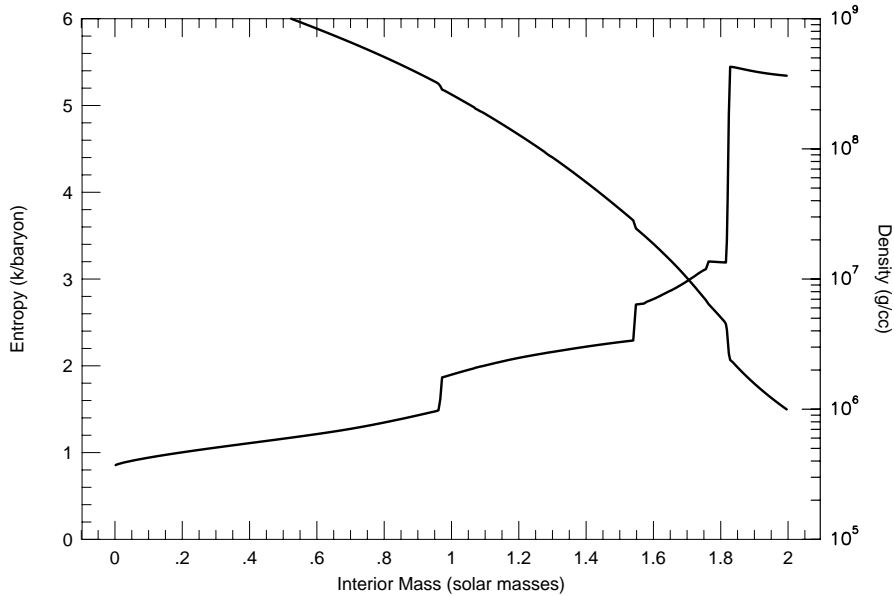,height=5in}}
\vskip -0.5 in
\caption{Entropy and density distributions inside a 20 solar mass
  presupernova star. The iron core mass here is 1.54 \Msun; the base
  of the oxygen shell is at 1.82 \Msun. The sudden decrease in density
  at the base of the oxygen shell causes an abrupt decline in ram
  pressure which often results in explosions happening with this mass
  cut.}
\lFig{fig2}
\end{figure}

The choices of piston mass and explosion energy are not free
parameters, but are highly constrained by observations. The piston
mass cannot be smaller than the iron core mass or unacceptable
overproductions of $^{54,58}$Fe and other neutron-rich species in the
iron group will occur. On the other hand it cannot be much larger than
the base of the oxygen shell ($S/N_Ak$ = 4) or, as we shall see,
typical neutron star masses will be too large. The large density
decrease associated with the base of the oxygen shell is also
dynamically important and successful explosion calculations, when they
occur, frequently find the mass cut there. The explosion energy is
constrained to be 1 - 2 B by observations of SN 1987A
\citep{Bet90,Arn89} which was a Type II supernova of typical mass
(about 18 - 20 \Msun). It is also constrained by the observed light
curves of Type II supernovae.

\subsection{Remnant Masses}
\lSect{remnants}

Observations by \cite{Tho99} of a large number of pulsars in binary
systems give a narrow spread in masses, $1.35 \pm 0.04$ \Msun.  There
must be room for some diversity, however. \cite{Ran05} present
compelling evidence for a pulsar in the Terzian 5 globular cluster
with a gravitational mass of 1.68 \Msun. The remnant gravitational
masses for our survey using the Kepler stellar evolution code, with KE
= 1.2 B and pistons located the edge of the iron core, are plotted in
\Fig{fig3}. A more careful analysis of fall back in these models using
an Eulerian hydrodynamics code and a better treatment of the inner
boundry conditions has been carried out by \citet{Zha07}, but gives
similar numbers for solar metallicity stars. Using the \citet{Zha07}
values, adopting a Salpeter initial mass function with $\Gamma = 1.35$
to describe the birth frequency of these stars, and assuming a maximum
neutron star mass of 2.0 \Msun, one obtains an average {\sl
gravitational mass} for the neutron star of $1.47 \pm 0.21$ if the piston
is at the $S/N_Ak$ = 4.0 point and $1.40 \pm 0.22$ if it is at the
edge of the iron core.  If the maximum neutron star mass is 1.7 \Msun,
the numbers are changed to $1.41 \pm 0.15$ \Msun \ and $1.34 \pm 0.14$
respectively.  In this paper, we carried out simulations with the
piston at both points - the iron core edge, and the base of the oxygen
shell.  Larger masses than typical are also possible for the rare
exceptionally massive star, usually those over 25 \Msun.  For those
cases where a black hole was made, its average mass was around 3
\Msun. We note that these numbers are for single stars and they could
be altered significantly in mass exchanging binaries.

\begin{figure}
\centerline{\epsfig{angle=90,file=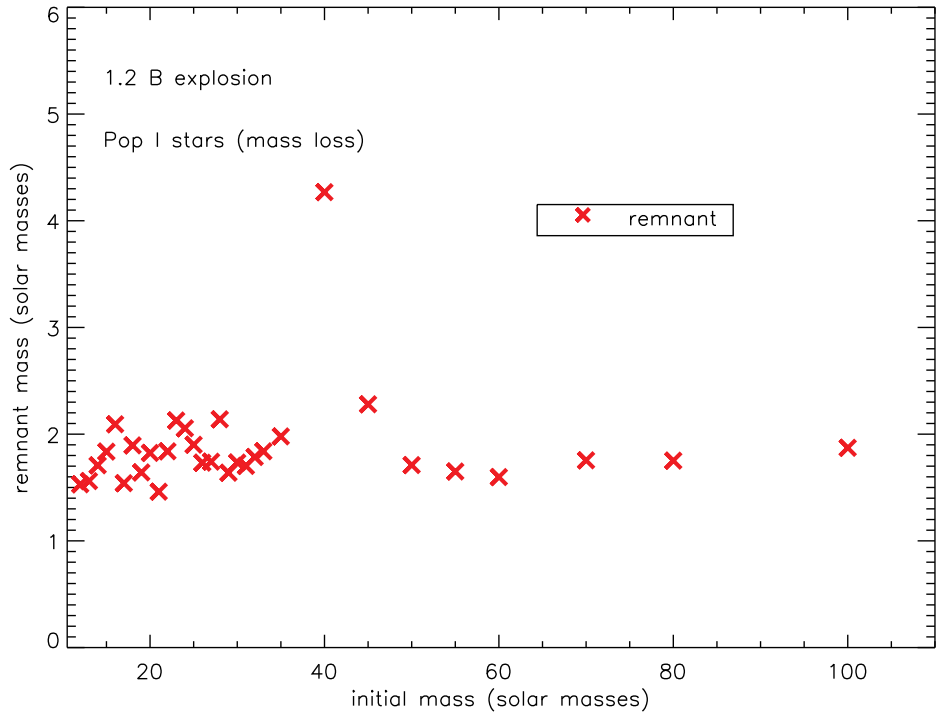,height=4in}}
\caption{Fe core masses for a grid of stellar masses. Solar
  metallicity stars. See text for explanation.}
\lFig{fig3}
\end{figure}

The figure also shows that neutron stars are made by both the lightest
main sequence stars and the heaviest. This is a consequence of mass
loss. The helium core mass of the presupernova star increases
monotonically with main sequence mass up to about 45 \Msun, where it
reaches a maximum of 13 \Msun. Beyond that the helium core shrinks due
to efficient Wolf-Rayet mass loss and the iron core mass shrinks with
it. A 100 \Msun \ model had a total mass of only 6.04 \Msun \ when it
died - all helium and heavy elements - and an iron core mass of 1.54
\Msun.

\begin{figure}
\centerline{\epsfig{angle=90,file=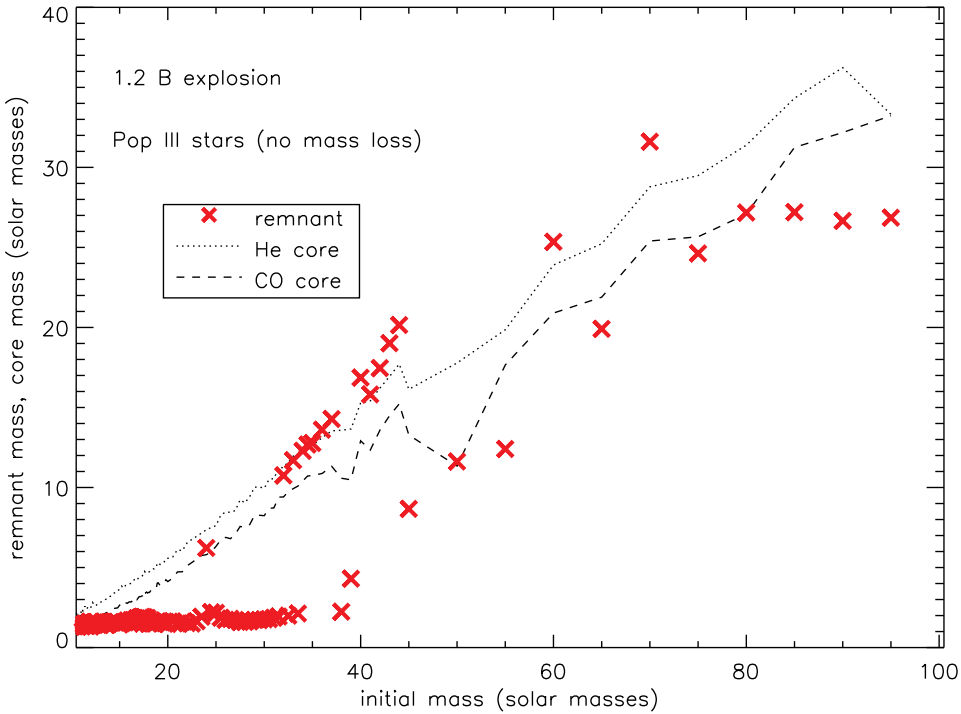,height=4in}}
\caption{Fe core masses for a grid of stellar masses. Zero metallicity
  stars. Many more black holes are made because the star loses little
  mass during its evolution to presupernova. The two branches of black
  hole masses at high main sequence mass correspond to red (lower
  branch) and blue (upper branch) supergiant progenitors. The lower
  carbon-oxygen core masses for the red supergiant cases reflect
  dredge up and primary nitrogen production \citep{Zha07,Heg07}.
\lFig{fig4}}
\end{figure}

The results are quite different for stars with low metallicity and,
hence, reduced mass loss \citep{Heg07,Zha07}. \Fig{fig4} shows that the
remnant mass increases rapidly for main sequence masses above about 35
\Msun \ and continues to increase at higher masses. These large masses
are due to fall back. A 1.2 B explosion is inadequate to unbind the
entire star, especially given the large helium core \citep{Woo02} and
effect of the reverse shock. A 100 \Msun \ main sequence star now dies
with a helium core of 42 \Msun, well into the pulsational pair
instability domain \citep{Heg02}. Unless supernova engines of much
greater power than 1.2 B become available at low metallicity, these
stars will make black holes, not neutron stars, and if the rotation
rate is sufficient, gamma-ray bursts.

One may also note the existence of two branches of black hole remnants
above 35 \Msun \ in \Fig{fig4}. \citet{Zha07} find that these branches
correspond to two different classes of progenitors - red supergiants,
which experience a lot less fall back during the reverse shock
\citep{Che89} - and more compact, extremely blue supergiants. If the
star produces primary nitrogen due to the interpenetration of the
helium convective core and hydrogen envelope, it swells to red giant
proportions, has a weaker reverse shock, and leaves a smaller remnant
mass.

\subsection{Light Curves}
\lSect{light}

The KEPLER code includes radiative diffusion and can thus be used to
calculate approximate light curves for the supernovae it produces. The
code is limited by using a single temperature for the radiation and
the matter, and assumes blackbody radiation, but these are not bad
approximations during the plateau stage of Type II supernovae
\citep{Wea80,Eas94}. The principal opacity source is electron scattering
with the free electron abundance determined by solving the Saha
abundances of all ions for the 19 isotopes in the reaction network
\citep{Ens88}. A floor opacity of 10$^{-5}$ cm$^2$ g$^{-1}$ is used in
regions that have recombined. The abundance of $^{56}$Ni is taken from
the nucleosynthesis calculation and moderate mixing of the helium core
is assumed.

The resulting light curves for four explosions of a 15 \Msun \
supernova are given in \Fig{fig5} for cases where the mass cut was
taken at the edge of the iron core and at the location where the
entropy equals 4$\,k_B$/baryon.  Two explosion energies, 1.2 B and 2.4
B were employed. The explosions that had the higher kinetic energy
were brighter on the plateau and the ones with the deeper mass cut
(and hence more $^{56}$Ni ejected) had the brighter tails. The mass of
$^{56}$Ni ejected was 0.086 \Msun \ for the 1.2 B explosion with the
mass cut at S = 4$\,k_B$/baryon; 0.096 \Msun for the 2.4 B explosion
with mass cut at S = 4$\,k_B$/baryon; 0.27 \Msun \ for the 1.2 B
explosion with the mass cut at the edge of the iron core; and 0.31
\Msun \ for the 2.4 B explosion with the mass cut at the edge of the
iron core.

Clearly, the models with higher kinetic energy are brighter on the
plateau \citep[see also][]{Pop93}. In fact, if the kinetic energy were
any larger than 2.4 B, the supernova would be far brighter than
average Type IIp supernovae. On the other hand if the explosion energy
were much less than 1 B, large amounts of material would fall back,
increasing the masses of the neutron star remnants beyond acceptable
values and robbing the nucleosynthesis of its most prolific sources.
We conclude that the range 1.2 - 2.4 B is the relevant one for modern
supernovae in solar metallicity stars and these are the values
employed in the nucleosynthesis survey.

\begin{figure}
\centerline{\epsfig{file=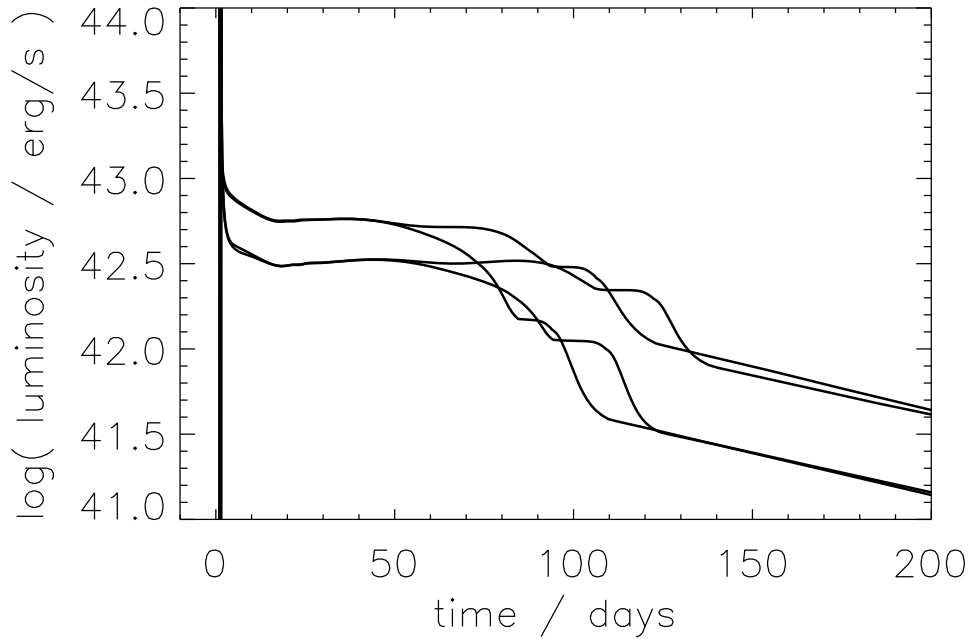,height=5.0in,angle=90}}
\caption{Light curve of a 15 solar mass supernova. The four curves
  represent two choices of explosion energy and combined with two
  choices of mass cut.  The presupernova star was a red supergiant.}
  \lFig{fig5}
\end{figure}

\section{Nucleosynthetic Yields}
\lSect{nucleo}

The integrated yields of the elements are given for four different
choices of mass cut and explosion energy in \Fig{fig6}. Whether one
places the piston at the edge of the iron core or the base of the
oxygen shell and whether the explosion energy is 1.2 B or 2.4 B makes
little difference except to the iron group. There the difference is of
order a factor of two, with lower iron yields obviously resulting from
lower explosion energies and shallower pistons. In all cases the iron
group synthesis is low compared both with C, O, Ne, and Na and with
$s$-process production above Ni. One expects from one-half to
two-thirds of the iron group to come from Type Ia supernovae
\citep{FXT95} which are not included here. The $s$-process, which is
secondary in nature, will be underproduced in stars of less than solar
metallicity, so a factor of two extra here relative to oxygen is not
undesirable.

\begin{figure}[t]
\centerline{\epsfig{file=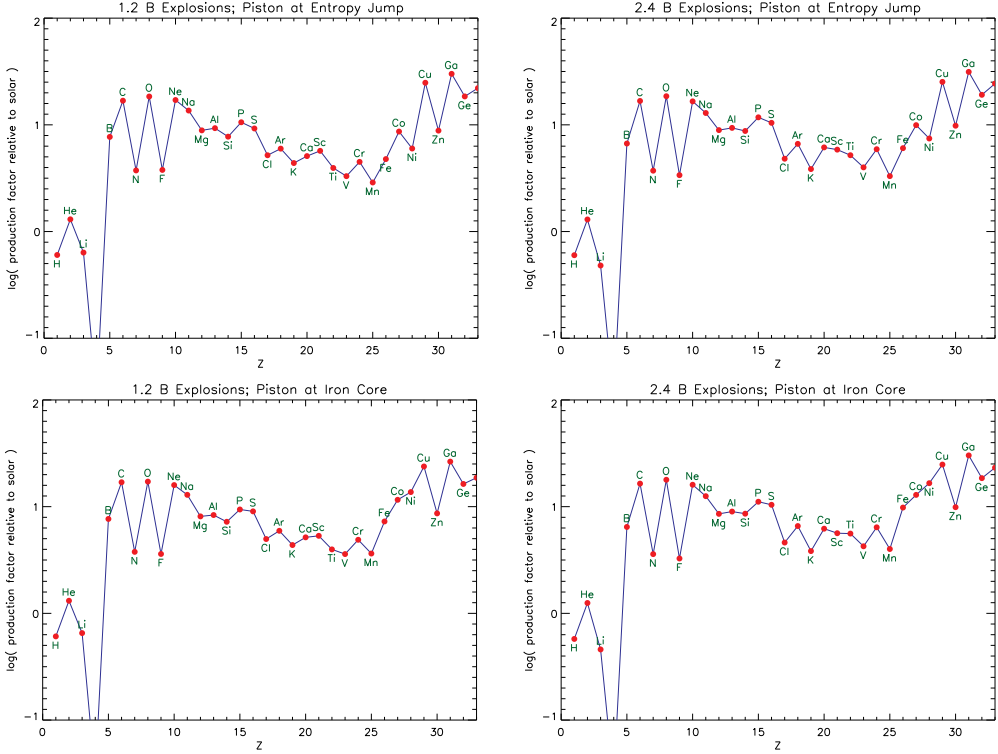,height=4.0in}}
\caption{Elemental yields integrated over a Salpeter initial mass
  function for solar metallicity stars with masses from 12 \Msun \ to
  120 \Msun.  Calculations were carried out for for two choices of
  explosion energy (1.2, 2.4 B) and two piston location (at the place
  where the entropy jumps to $S/N_Ak = 4$ and at the edge of the iron
  core). Only minor differences are discernible. Similarly small
  changes occur when the slope of the IMF is changed from -1.35
  (Salpeter) to a gentler -0.9.}
\lFig{fig6}
\end{figure}

\Fig{fig7} shows the integrated nucleosynthesis (the yields folded
with a Salpeter initial mass function with $\Gamma$ = -1.35) for all
elements up to lead compared with the isotopic composition of the
sun. \Fig{fig8} shows the corresponding comparison of
isotopes. Overall, the agreement is quite good, especially considering
that several known sites of important nucleosynthesis have been
omitted. Classical novae will need to produce $^{15}$N and $^{17}$O,
though some $^{15}$N is made here by the neutrino spallation of
$^{16}$O. The isotopes $^{44}$Ca, $^{47}$Ti, and $^{48}$Ca are
underproduced and may come from some rare form of SN Ia \citep{Woo97}
or asymmetric supernova. The overabundance of $^{40}$K is not a
concern since some will decay before the sun is born. Also missing are
ordinary Type Ia supernovae, which contribute half or more of the iron
group, and asymptotic giant branch stars which make $^{14}$N and the
$s$-process. In fact, the full production of carbon here is a novel
and surprising result, since it is usually attributed to low mass
stars.  It is made here chiefly in the winds of very massive
Wolf-Rayet stars and its production is facilitated by the new lower
solar abundance \citep{Lod03}.

\begin{figure}
\centerline{\epsfig{file=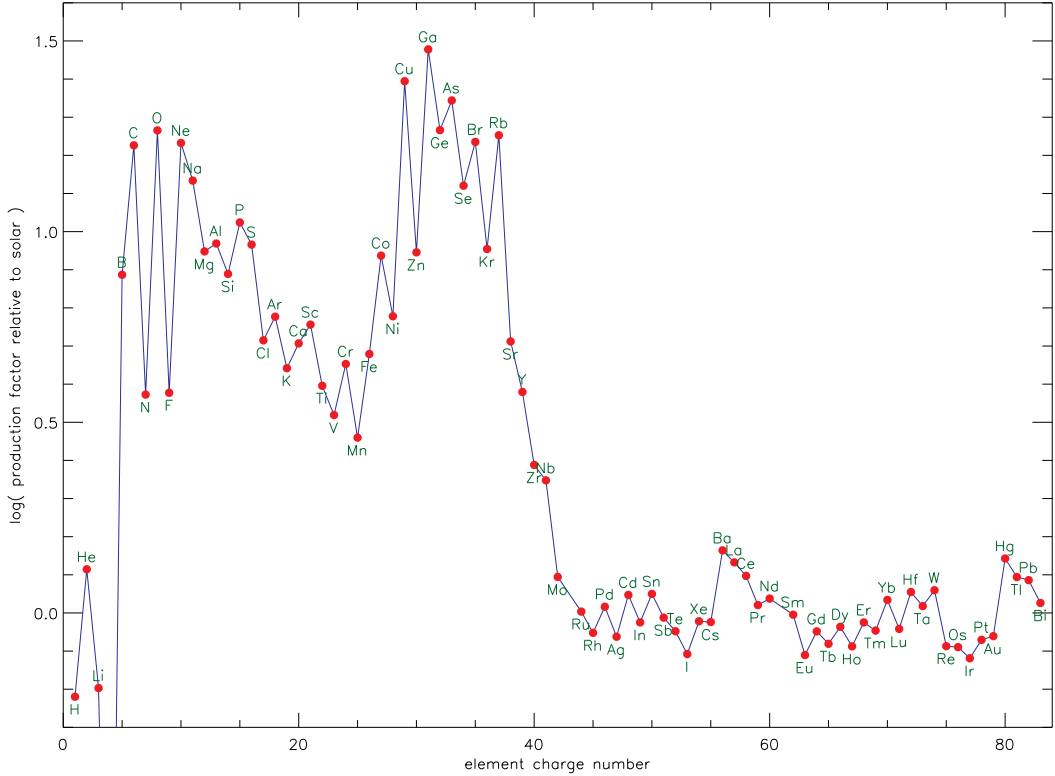,height=5.5in,angle=90}}
\caption{Elemental yields when the grid of supernova masses is
  integrated over a Salpeter initial mass function. The explosion
  energy was 1.2 B and the piston was located at the mass coordinate
  where $S/N_Ak$ = 4.0. A strong s-process operates up to Z = 40.}
  \lFig{fig7}
\end{figure}

Finally conspicuously absent is the $r$-process and other products of
the neutrino-powered wind. The wind of a young neutron star is a
prolific source of heavy elements, accounting for about half of the
isotopes in nature. These include not only the $r$-process
\citep{Woo94}, but some important $p$-process nuclei
\citep{Pru06,Fro06}, and even abundant elements like Zn \citep{Hof96}
  and Sc \citep{Pru05}. Hans Bethe was very interested in the
  neutrino-powered wind and the $r$-process, and he was working on it
  when SEW last saw him in Winter 2003.  This was probably his
  last supernova-related project. He said that he had an abiding
  interest in uranium.

\begin{figure}
\centerline{\epsfig{file=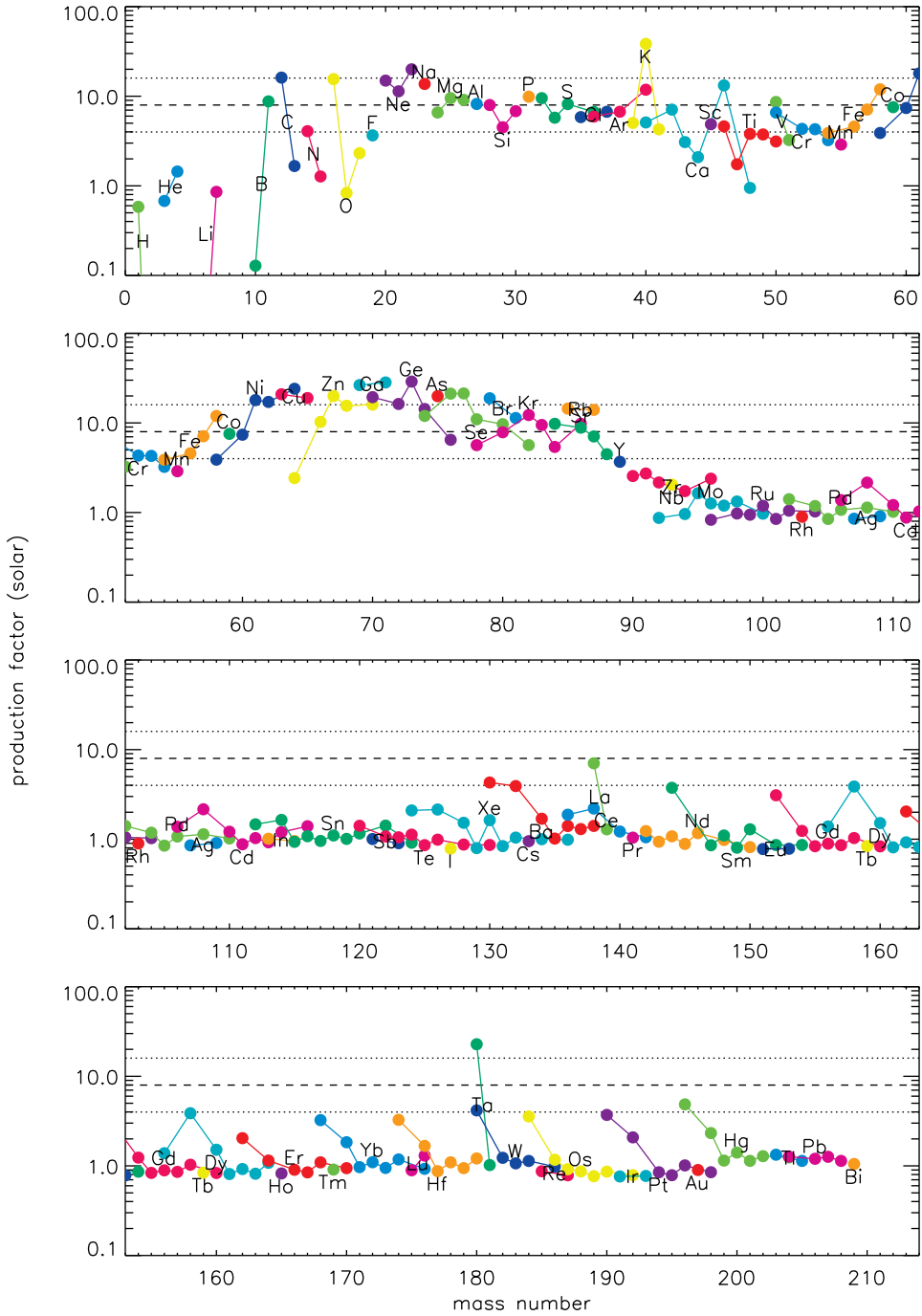,height=5in,angle=90}}
\caption{Isotopic nucleosynthesis integrated over a Salpeter initial
  mass function. The results are in good agreement with solar
  abundances below A = 85. The excess of $s$-process nuclei above the
  iron group is needed in part, to compensate for the smaller
  production of these secondary nuclei in stars of slightly lower
  metallicity. The $\gamma$-process is mostly successful in making the
  neutron-deficient isotopes (the ``p-nuclei'') above A = 130, but
  there is an annoying deficiency of p-nuclei production from A = 90
  to 130. The large productions of $^{11}$B, $^{19}$F, $^{138}$La,
  and $^{180}$Ta, are due to the neutrino process. Products of the
  neutrino wind, e.g., $^{64}$Zn and the $r$-process, are not included
  in this plot.}
\lFig{fig8}
\end{figure}

\section{The Special Cases of $^{26}$Al and $^{60}$Fe}
\lSect{fe60}

Having computed the isotopic nucleosynthesis in a grid of stars up to
120 \Msun, including the contribution of the winds the more massive
stars make as Wolf-Rayet stars, we turn to the examination of two
isotopes of special interest to gamma-ray line astronomy. The
long-lived isotopes $^{26}$Al and $^{60}$Fe accumulate in the
interstellar medium from thousands of supernovae and thus serve as
calibrations on the integrated yields of massive stars. Observations
by RHESSI \citep{Smi04} and INTEGRAL \citep{Har05} give a ratio of
fluxes from the decays of $^{26}$Al and $^{60}$Fe of 0.16 and 0.11
$\pm$ 0.03 respectively. Both measurements are quite consistent with
the predicted value, 0.16, by \citet{Tim95} based upon yields from the
\citet{WW95} survey. Later calculations \citep{Rau02,Lim03}, however,
using stellar and nuclear physics that was nominally ``improved'' gave
a much larger synthesis of $^{60}$Fe that was not in line with
observations \citep{Pra04}.

The ratio of gamma-line fluxes implies, in steady state, a synthesis
ratio by mass of $^{60}$Fe/$^{26}$Al of 60/26 times the flux ratio, or
about 0.3.  (The steady state abundance is inversely proportional to
the decay rate and the flux is the abundance times the decay rate so
the decay rate itself cancels.) \citet{Tim95} gave a theoretical ratio
of 0.38 with an expected uncertainty of a factor of 1.7. Using the
larger grid of models here, however, and including mass loss as
discussed in \Sect{mdot}, we calculate a ratio of 1.8, i.e., six times
too large. This large excess of $^{60}$Fe/$^{26}$Al is consistent with
what \citet{Rau02} found, even though their calculations did not
include the quite massive stars studied here (up to 120 \Msun), nor
their mass loss. What is wrong?

\subsection{Nuclear Physics Uncertainties}
\lSect{nuc60}

One problem is certainly the use of uncertain nuclear reaction rates
in all studies to date. In making the transition to the reaction rate
data base of \citet{Rau00}, we erroneously used the new Hauser
Feshbach rates especially for $^{26}$Al(n,p)$^{26}$Mg and
$^{26}$Al(n,$\alpha$)$^{23}$Na. These are the principal means of
$^{26}$Al destruction in the carbon and neon layers where $^{26}$Al is
explosively synthesized. The Rauscher-Thielemann rate for
$^{26}$Al(n,p)$^{26}$Al at $2 \times 10^9$ K, for example, is $1.4
\times 10^8$ cm$^3$ Mole$^{-1}$ s$^{-1}$. The Rauscher-Thielemann rate
for $^{26}$Al(n,$\alpha$)$^{23}$Na at $2 \times 10^9$ K is $2.6 \times
10^7$ cm$^3$ Mole$^{-1}$ s$^{-1}$. These are a both a factor of 3 to 5
higher than the rates used for these reactions by \citet{WW95} and the
{\sl experimental} determinations by \citet{Koe97} and \citep{Cau88}.

A second effect, less important than the cross sections, is a
super-hot helium shell ($4 \times 10^8$ K) in several of the
pre-supernova star. This shells existence was traced to the use of
OPAL opacities in a region where they may not be appropriate, a region
where electron scattering dominates. Using the electron scattering
opacity of \citet{Wea78} just in high temperature regions where
electron scattering dominates decreased the $^{60}$Fe yield
significantly, but this was only in a few stars.

Using what we believe to be more nearly correct cross sections for
$^{26}$Al destruction (though still uncertain) and adjusting the
opacity as described, the integrated yield of $^{60}$Fe to $^{26}$Al
is reduced to 0.95. This is for a standard Salpeter IMF with $\Gamma$ =
-1.35. If we instead change the slope to -0.90, i.e., enhance the
production of very massive stars, the ratio is reduced slightly to
0.81. Even then one-half the yield of $^{26}$Al comes from stars under
35 \Msun, not the more massive ones and their Wolf-Rayet winds.

There are further uncertainties to explore, however. Several of the
cross sections governing the production of $^{60}$Fe are also highly
uncertain. The reaction $^{59}$Fe(n,$\gamma$)$^{60}$Fe affects its
synthesis and $^{60}$Fe(n,$\gamma$)$^{61}$Fe controls its
destruction. Neither is measured, though both could be, admittedly
with difficulty. Interestingly, both changed in the \citet{Rau00}
tabulation in such a direction as to increase $^{60}$Fe
production. The tabulation by \citet{Woo78} had, for helium burning
temperatures, a rate for $^{59}$Fe(n,$\gamma$)$^{60}$Fe half as large
and a rate for $^{60}$Fe(n,$\gamma$)$^{61}$Fe twice as large. When the
older rates were used for a select set of models, the $^{60}$Fe
production was reduced by about a factor of two.

The final {\sl nuclear} uncertainty is the rate governing the
production of neutrons where $^{60}$Fe is made, i.e.,
$^{22}$Ne($\alpha$,n)$^{25}$Mg. The rate included in our network
\citep{Jae01} is increased from what was used in 1995. If we reduce
its value in a few select models by a factor of two (within the error
bar), $^{60}$Fe production is again decreased by up to a factor of
two, though usually the effect was smaller.

All things considered, variation of only the nuclear physics, bringing
uncertain cross sections back to the values they had in the Timmes et
al survey, could account for most of the difference in the present
calculations and the observations. \emph{Hence further progress in
this important field of astronomy depends upon more accurate
measurements and estimates of critical nuclear physics.}

\subsection{Uncertainties in the Stellar Models}

This is not to say that non-nuclear effects are
unimportant. Metallicity, mass loss, rotation, and an uncertain IMF
certainly all play major roles. \citet{Pal05} have explored $^{26}$Al
production in models of massive stars that include rotationally
induced mixing, as well as mass loss and different choices of
metallicity. An explicit comparison with a couple of our models is
educational. For a 60 \Msun \ main sequence star with Z = 0.02 and no
rotation, they find an $^{26}$Al production {\sl in the pre-explosive
  wind of the star} of $1.30 \times 10^{-4}$ and a final star mass of
12.4 \Msun \ \citep{Mey00}. For our 60 \Msun \ model with metallicity Z
= 0.016 and using the smaller experimental rates for
$^{26}$Al(n,p)$^{26}$Mg and $^{26}$Al(n,$\alpha)^{23}$Na, we find a
production in the wind of $1.1 \times 10^{-4}$ \Msun \ and a final
mass of 8.0 \Msun. But we also find an additional $9.9 \times 10^{-5}$
\Msun \ of $^{26}$Al is produced {\sl in the explosion} of the
remaining star, chiefly by explosive neon burning. This is good
agreement, and shows that the explosion and wind may contribute
comparable amounts to $^{26}$Al synthesis even for a 60 \Msun
\ progenitor. \citet{Pal05} further explore the dependence of
metallicity and rotation though, and find $^{26}$Al production in the
wind of this same star is increased to $2.2 \times 10^{-4}$ \Msun \ if
the rotation rate is 300 km s$^{-1}$ on the main sequence, or $3.0
\times 10^{-4}$ \Msun \ with no rotation but Z = 0.04. Combining both
effects, Z = 0.04 {\sl and} $v_{\rm rot}$ = 300 km s$^{-1}$, the
$^{26}$Al production in the wind becomes even larger $7.2 \times
10^{-4}$ \Msun. While one must be concerned that increasing the
metallicity may also increase the $^{60}$Fe yield and thus not
increase the $^{60}$Fe/$^{26}$Al ratio \citep{Pra04}, this does show
the sensitivity of $^{26}$Al to reasonable variations in rotation
rate. For a 120 \Msun \ model, the effect is even greater is similar.
For Z = 0, $v_{\rm rot}$ = 0, \citet{Pal05} obtain an $^{26}$Al mass
of $5.7 \times 10^{-4}$ \Msun \ in the wind while we have $4.9 \times
10^{-4}$ \Msun \ plus $2.9 \times 10^{-5}$ \Msun \ made in the
explosion. With Z = 0.04 and $V_{\rm rot}$ = 300 km s$^{-1}$,
\citet{Pal05} get a whopping $2.2 \times 10^{-3}$ \Msun.

\citet{Lim06} have also recently (after our present study was
completed) examined the sensitivity of $^{60}$Fe and $^{26}$Al
production to the prescription for mass loss and the slope of the
IMF. They find that both can make an appreciable difference.

\section{Conclusions}
\lSect{concl}

We still don't understand exactly how massive stars explode, far less
the variation of explosion properties - especially mass cut and
explosive kinetic energy - with main sequence mass. This remains a
forefront problem in nuclear astrophysics research to which Hans Bethe
contributed greatly. It is likely, in the final analysis, that the
physical intuition, terminology, and convective, neutrino-powered
paradigm that he and his colleagues brought to the field will form the
basis of a complete understanding, though we aren't there yet
\citep{Woo05}. Certainly, the low entropy, super-nuclear density
bounce following the initial collapse will be phase one of any massive
star explosion.

This lack of a first principles understanding of the explosion
mechanism, however, is not a fundamental roadblock on our path to
understanding the origin of (almost all of) the elements. Arguments
have been presented here to show that the mass cut is highly
constrained by nucleosynthesis and observed neutron star masses. The
explosion energy in common Type II supernovae is also mostly in the
range 1.2 B plus or minus a factor of two. Exploding a large range of
stellar masses with pistons located either at the edge of the iron
core or the base of the oxygen burning shell - the maximum range
allowed - and with explosion energies of either 1.2 B or 2.4 B gives
very similar nucleosynthesis. The iron group is most affected and the
magnitude of the uncertainty is about a factor of two.

The nucleosynthesis that results (\Fig{fig7} and \Fig{fig8}) agrees
reasonably well with solar abundances. There are some changes caused
by the recent downward revisions of the solar CNO abundances, and at
first glance the agreement is worsened by these changes. Since
$^{16}$O is our standard normalization point in nucleosynthesis
studies, since we now need to make less of it, the production of all
other heavier elements is decreased. Yields that previously would have
coproduced Si and O say, in solar proportions, now overproduce O
(\Fig{fig8}). The production of odd-Z elements and odd-A isotopes is
also decreased because the initial CNO in the star later becomes the
$^{22}$Ne that sets the neutron excess for carbon, neon, and oxygen
burning (\Sect{solabn}). Still the agreement is not too bad, and most
of the missing species $^{13}$C, $^{14,15}$N, $^{48}$Ca, etc. can be
attributed to other sites than massive stars.

The outstanding problem in nucleosynthesis theory presently is a full
understanding of the $r$- and $p-$processes. The latter has an
appreciable contribution from explosive neon and oxygen burning (shown
in \Fig{fig8}) for A greater than 130, but is underproduced for
lighter masses.  The solution for both the $r$-process and the light
$p$-process probably lies in the neutrino-powered wind. Current models
give inadequate entropy in the wind and this may be where
nucleosynthesis can be an important diagnostic of the explosion model
\citep[e.g.,][]{Bur06}.

The nucleosynthesis of the long-lived radioactivities $^{26}$Al and
$^{60}$Fe is an important constraint on the stellar models, and one
that is largely independent of the explosion mechanism. The abundances
inferred from gamma-ray line astronomy may have important implications
for rotationally induced mixing, convection theory, mass loss theory,
the initial mass function for massive stars, and the distribution of
metals in the galaxy. The synthesis is also quite sensitive to nuclear
reaction rates whose uncertain values could be better determined in
the laboratory, however.  In particular, the discrepancy between
observations of the $^{60}$Fe/$^{26}$Al ratio and recent calculations
- this work and \citet{Rau02} - may involve a ``perfect nuclear
storm'' of erroneous choices. The rates affecting $^{26}$Al
destruction were almost certainly too high; the rates affecting
$^{60}$Fe production, namely $^{59}$Fe(n,$\gamma$)$^{60}$Fe and
$^{22}$Ne($\alpha$,n)$^{25}$Mg, may have been too high; and the rate
for its destruction, $^{60}$Fe(n,$\gamma$)$^{61}$Fe may have been too
low. Given the choices made by \citet{WW95}, the prediction of
\citet{Tim95}, which agrees with observations, is still defensible.
In any case, important inferences about the stellar models will only
be credible (and necessary), when these uncertain rates have been
better determined.


\section*{Acknowledgments}

This research reported here formed the basis of the 2005 Bethe Prize
lecture presented by SEW at the April, 2005 meeting of the American
Physical Society. The work has been supported by the NSF (AST 0206111)
and the DOE SciDAC Program (DOE DE-FC-02-01ER41176 and DOE
DE-FC-02-06ER41438).  AH was supported under the auspices of the
National Nuclear Security Administration of the U.S. Department of
Energy at Los Alamos National Laboratory under Contract
No. DE-AC52-06NA25396.  The authors are grateful to Rob Hoffman for
assistance with the nuclear reaction data base.


\end{document}